# What is in a Name: Defining "High Entropy" Oxides

Matthew Brahlek[1], Maria Gazda[2], Veerle Keppens[3], Alessandro R. Mazza[4], Scott J. McCormack[5], Aleksandra Mielewczyk-Gryń[2], Brianna Musico[6], Katharine Page[3], Christina Rost[7], Susan B. Sinnott[8], Cormac Toher[9], Thomas Z. Ward[1,a)], Ayako Yamamoto[10]

[1]Materials Science and Technology Division, Oak Ridge National Laboratory, Oak Ridge, TN
[2]Institute of Nanotechnology and Materials Engineering, and Advanced Materials Centre, Gdańsk University of Technology, Gdańsk, Poland
[3]Department of Materials Science and Engineering, University of Tennessee Knoxville, Knoxville, TN
[4]Center for Integrated Nanotechnologies, Los Alamos National Laboratory, Los Alamos, NM
[5]Department of Materials Science and Engineering, University of California Davis, Davis, CA
[6]Sigma Division, Los Alamos National Laboratory, Los Alamos, NM
[7]Department of Physics and Astronomy, James Madison University, Harrisonburg, VA
[8]Department of Materials Science and Engineering, Pennsylvania State University, University Park, PA
[9]Department of Materials Science and Engineering and Department of Chemistry and Biochemistry, University of Texas at Dallas, Richardson, TX
[10]Graduate School of Science & Technology, Shibaura Institute of Technology, Tokyo, Japan

[a)]Author to whom correspondence should be addressed: wardtz@ornl.gov

*Abstract*

High entropy oxides are emerging as an exciting new avenue to design highly tailored functional behaviors that have no traditional counterparts. Study and application of these materials are bringing together scientists and engineers from physics, chemistry, and materials science. This diversity of disciplines each comes with perspectives and jargon that may be confusing to those outside of the individual fields, which can result in miscommunication of important aspects of research. In this perspective, we provide examples of research and characterization taken from these different fields to provide a framework for classifying the differences between compositionally complex oxides, high entropy oxides, and entropy stabilized oxides, which is intended to bring a common language to this emerging area. We highlight the critical importance of understanding a materials crystallinity, composition, and mixing length scales in determining its true definition.

*Introduction*

The functional properties of metal oxides constitute one of the widest research fields belonging to many fundamental and applied scientific disciplines e.g. materials science, chemistry, physics, engineering and many others. Simplifying this, there are two main conventional ways of tailoring the functional properties of oxides. The first one is a chemical modification through substitution of the host cations with others chosen based on their electronic configuration, ionic radius, and other basic features. The second way of influencing material properties is choosing the most appropriate morphology and micro- or nanostructure, which may be realized using various synthesis techniques. For simplicity, controlling materials relies on what elements are in the material and how those elements' nearest neighbor environments influence electron interactions. Designing and realizing a material having desired functional properties is a difficult task since the chemical composition, morphology, microstructure, stability, functional properties, and synthesis are not independent. Furthermore, the thermodynamic stability during synthesis may limit the ability to realize these functionalities in real materials.

Utilizing configurational entropy as the thermodynamic driver for stable solid solutions allows access to a vast and unexplored materials space that is unattainable with enthalpy stabilized methodologies. Ideal entropy can be considered as the value of the entropy in the limit as the system becomes fully disordered – as ordering increase, the system's entropy value reduces away from this ideal value. Conceptually, a system's entropy might be viewed as a bank, in which entropy can either be deposited to randomize structure, or withdrawn to stabilize materials with chemical, structural, or charge ordering within the high-entropy matrix. The key to utilizing high entropy materials for desirable applications is the ability to exploit novel functional structure-property relationships by manipulating entropy to form unique crystal structures, valence states, coordination numbers, and local distortions in the lattice. In essence, the entropy term might be considered as a controllable order parameter which can provide a route to designing highly tunable properties and novel emergent functionalities. The number of variables influencing functional properties in this field is enormous. The uniqueness of this wide group of materials is that they show a long-range structural crystalline order and short-range compositional disorder. The short range disorder which arises from the compositional complexity influences functional properties of materials such as their thermal and thermoelectric[1], dielectric[2], magnetic[3,4], and many other properties. These depend on the valence states, electronegativities, preferred coordination numbers, ionic radii, and masses of particular cations and is a feature which may be controlled only in a very limited way. Moreover, the short-range disorder competes with short-range order, as it is observed both in high entropy metal-metal bonded alloys[5] and ionically and covalently bonded oxides[6]. This point is a particular point of contention as we explore the paradigms of defining the "level of entropy" in a given composition both from a theoretical and an experimental point of view.

The choice of the particular way to evaluate if a given material should be considered a "high entropy" oxide is troublesome. Many published reports neglect to calculate configurational entropy to determine if a particular material should be classified thermodynamically as a high entropy oxide, with the moniker often applied to any material hosting 5 or more unimolar components. The issue of entropy metrics is rarely addressed, which makes it hard to compare the influence of configurational entropy on specified properties. Another issue is how to differentiate between materials with different configurational entropy. For example, some materials which are classified as "high entropy" oxides with metrics based solely on entropy values resulting from the single-lattice model are "medium entropy" oxides when one accounts for entropy metrics developed based on the sublattice model. In the literature, the nomenclature of these materials is not clearly defined which results in some discrepancies between the most common terms,

"compositionally complex", "high entropy", or "entropy stabilized". In this perspective, a hierarchy of terminology is outlined in an attempt to address these discrepancies. Emphasis is placed on oxides, but the discussion is meant to hold for all materials where cation and anion sublattice are simultaneously present. Other high-entropy ceramics with anion sublattices include carbides, nitrides, borides, halides and silicides[7]; and multi-anion high-entropy systems are relevant to these discussions.

We propose that materials having multiple elements residing on one or more sublattice can be considered as "compositionally complex" materials (CCM). "High entropy" oxides (HEO) are defined as materials where the configurational entropy plays some role in formation and stability but is not necessarily dominant over enthalpy; these materials may include native defects in place of having multiple elements which can influence stability. "Entropy stabilized" oxides (ESO) are then a subset of high entropy materials with the strictest definition, in which entropy dominates the enthalpy term and is critical in the formation of a crystalline phase in which one or more of the constituent parent materials may not exhibit. This hierarchy is presented in Figure 1. The entropy stabilized title does not necessarily need to be restricted to materials where equilibrium entropy stabilization persists to room temperature. As an example, some applications have higher operating temperatures that may be sufficiently high to maintain entropy's dominant role in maintaining stable mixing in a material, however that material my begin to destabilize at room temperature. There is nothing intrinsically special about room temperature. Rather, it seems appropriate for there to be a broader criterion where entropy stabilization requires only that material evolution is substantially entropy-influenced at some point in the synthesis history, and that the structure and properties of the highly disordered state can be studied and functionally implemented. The outline of this perspective is organized as follows: A brief overview providing specific examples of importance of compositionally complex materials in relation to different fields is provided. This is followed by a theoretical perspective on how to define a high entropy oxide system in relation to entropy stabilization and high, medium, and low entropy. The following section provides an overview of different synthesis methods being exploited to create these materials and highlights the fact that specific synthesis methods result in inherently different thermodynamics which drive inherently different crystallinity, compositions, and mixing length scales. The importance of multi-scale characterization is next described; since without clear understanding of a specific materials crystallinity, composition, and mixing length scales, it is impossible to apply the theoretical definitions. In the final section, an outlook towards the future of materials research in the "high entropy" arena is explored.

*The Connection of Chemical Complexity to Functionality*

Compositional complexity may lead to high configurational entropy, which, at high temperature, contributes to achieving a minimum Gibbs free energy, and in entropy stabilized oxides may be the driving force of the formation of a single-phase compound. High configurational entropy alone does not necessarily drive functional properties. Instead, it is more correct to recognize that the presence of compositional complexity and local interactions resulting from short-range disorder determine material properties. High configurational entropy is then just a measure of degree to this disorder. In oxides and other ceramics, owing to the chemically uniform anion sublattice, the local disorder is lower than that in alloys. Assuming that the *M* site of a *M*O oxide is occupied by five or more cations, the disorder appears in the O-*M*-O-*M* bonds and manifests itself as a distribution of the *M* cation coordination numbers, as well as the O-*M* bond lengths, degree of covalence, energies, local bond angles, vibration frequencies, *etc*. All these influence functional properties, especially those which are related either to the nearest and next nearest neighbor interactions (e.g. magnetic properties) or to the charge, energy and mass transport phenomena such as electronic, ionic, thermal conductivity, catalytic activity and thermoelectric effects. On the other hand, high-temperature stability as well as other high-temperature properties of ESOs inevitably are influenced by configurational entropy. Functionality is caused by the local microstate composition which can be correlated to configurational entropy. From this, there are two main drivers influencing CCM behaviors: the elemental composition and the degree of mixing. To highlight this point, it is illustrative to put this into specific contexts of where and how composition and mixing influence functionality.

Ionic conduction relies on thermally activated ion transport that in solids usually occurs as hopping of mobile ions through either vacancies or interstitial positions. The degree of compositional complexity can have a profound impact on functionality. The type of mobile charge carriers and their concentrations are a direct result of elements present within the system. For instance, the number of acceptor-type $B_i$ constituents in a $Ba(B_1,B_2,..,B_N)O_3$ perovskite determines the concentration of extrinsic oxygen vacancies. If the conduction occurs via oxygen vacancies, the mobile charge carrier concentration is the number of vacancies able to exchange sites with oxide ions which, in the simplest case, equals the concentration of mobile oxide ions. Additionally, the unit cell symmetry and parameters, lattice free volume, and the average distance between the consecutive hopping sites influence mobility. Generally speaking, a low hopping distance and high crystal lattice symmetry enhance ion mobility by decreasing the activation energy. On the other hand, the correlation between the lattice free volume and the activation energy of mobility depends on the particular material. For instance, the activation energy of oxygen ion transport in perovskite cerates and zirconates decreases while it increases in fluorite zirconia, ceria, and thoria with increasing lattice free volume[8]. Since ionic conduction is thermally activated, atom vibration frequencies and lattice stiffness affect the activation energy of conduction. In the case of proton and oxide ion conduction in *AB*O$_3$ perovskites, the most important are the O-*B*-O bending vibrations[9]. In the case of oxides where the *B* sites are occupied by multiple elements differing in atomic mass and characteristics of bonds with oxygen, the frequency of these vibrations depends on the chosen set of *B* cations.

While compositional complexity influences both the preexponential term and activation energy of conduction, the degree of short-range mixing and disorder can influence both these values as well. Let us consider the doped ceria ($Ce_{0.8}R_{0.2}O_{2-\delta}$, where R is La, Nd, Sm, Gd, Y, and Yb) as an example of an oxide ion conductor. Yashima and Takizawa reported that in all doped oxides the static and dynamic positional disorder of both cations and anions are higher compared with that in $CeO_2$[10]. The larger dynamic disorder of oxide ions was found to increase ionic conductivity of $Ce_{0.8}R_{0.2}O_{2-\delta}$. On the other hand, static disorder introduces a distribution of energy barriers for oxide ion transport. As a result, the activation energy of conduction increases while the number of equivalent conduction paths decreases. Another effect of static short-range disorder is that defects as well as ions can occupy sites of different energies. Therefore, they may be trapped at the low-energy sites, leading to an increase of activation energy of conduction. In contrast to oxide ion conducting oxides, the influence of compositional complexity and short-range disorder of lithium conducting oxides was completely different, which can give rise to unexpected functional improvements over less complex systems—as example, the very high ionic conductivity in $(MgCoNiZn)_{1-x-y}Ga_yLi_xO$ containing more than x=0.2 lithium ions[11]. Interestingly, the ionic conduction is proposed to occur through the oxygen vacancies, however a detailed description of the conduction mechanism requires further studies as other works have demonstrated the importance of Li vacancies in the conduction mechanism as well[12].

Similar to ion conduction, electronic conduction of compositionally complex materials depends on many subtle features related both to the constituent elements present and to the details of the local microstates governed by degree of mixing. As a result, the electrical properties may be controlled by microscopic properties, like oxidation state, spin state, defect types and concentration, and/or by external conditions, such as temperature and magnetic fields. In thermally activated electronic conduction in oxides, the composition determines both the crystal structure and band structure, so that it also controls the type and concentration of electronic charge carriers and mechanism of transport. In many oxides, the conduction mechanism is electron/small polaron hopping between localized states. If this is the case, the compositional complexity and short-range disorder may be expected to influence the preexponential factor and activation energy of conductivity in a very similar way to that described above for ionic conduction. For example, Shi *et al.* in the series of oxides based on $A$MnO$_3$ in which the $A$ site was substituted with five cations (A= La, Nd, Sm, Pr, Y, Gd, Ba, Ca and Sr) found that the oxide with the highest value and the lowest activation energy of conductivity contained 40% of acceptor constituents and showed the lowest cation size difference[13]. One of the examples of compositionally complex oxides exhibiting the band-type conduction is the $(Gd_{0.2}Nd_{0.2}La_{0.2}Sm_{0.2}Y_{0.2})CoO_3$ perovskite which may be considered as a semiconductor with a low-temperature band gap of about 1.2 eV while above 240 K it is 0.2 eV [14]. In this material, the composition, especially the magnetic state of the $A$ cations controls the electric properties of the material. The third group of CCM electronic conductors are those in which the presence of short-range disorder leads to the electron localization, as postulated in $Sr(Ti_{0.2}Fe_{0.2}Mo_{0.2}Nb_{0.2}Cr_{0.2})O_3$[13]. While less well-studied, the inherent electron correlations and diversity of coupled disordered microstates and related systems can also be expected to contribute to emerging quantum materials applications.

Instead of long-range disorder and short-range order, high entropy materials are structurally well-ordered in the long-range and chemically disordered in the short-range. In creating short range chemical disorder, the local spin configuration and magnetic interactions[15–18], charge configuration[19–21], electronic structure[22–25], and lattice[26–29] can be cultivated towards a desired phenomenon. This has recently been demonstrated in the case of correlated magnetism[16], where a prediction based on first nearest neighbor exchange interactions could be used to determine and then tune the magnetic interaction type and order temperature in HEOs. Further, this has been applied to several other collective phenomena including electronic conductivity[22,25,30], ferroelectricity[31], and phonon excitations[32].

Complexity can give rise to emergent electronic and magnetic phases that have laid the foundations for new technologies such as high-temperature superconductors, magnetic spin liquids, and topological materials. The goal to functionalize and ultimately utilize the properties of these so-called quantum materials requires understanding the key ingredients that stabilize these phases. Metals and insulators often exist in chemically simple, crystalline systems with high levels of purity and low levels of defects. However, many quantum phases only emerge in systems with large levels of doping or chemically complex compounds. Key examples are high temperature superconductors, $Bi_2Sr_2Ca_2Cu_3O_{10}$, metal-insulator transition materials $La_{1-x}Sr_xMnO_3$, charge compensated topological materials $A_y(Bi,Sb)_2(Te,Se,S)_3$, where $A$ can be Cr or V for magnetic doping or a variety of charge dopants. Thus, they necessarily coexist with high levels of disorder, where there has been realization that these effects cannot be ignored and may play a dominant role in stabilizing novel quantum phases. This suggests that careful control over entropy via chemical complexity may allow the previous desire to minimize disorder to be flipped, where disorder can be used to design and manipulate quantum-based phenomena through the design of novel many body systems.

*Theoretical Quantification of Material Entropy*

The total configurational entropy for statistically ideal mixing in an arbitrary crystal, is defined by Eq. 1:

(1) $S_{config} = -R \sum_j m_j \sum_i X_{ij} \ln X_{ij}$

where R is the gas constant, $m_j$ is the atoms per formula unit on the jth crystallographic site (site multiplicity multiplied by the formula unit) and $X_{ij}$ is the mol fraction of species $i$ on the $j$th crystallographic site. For high symmetry metallic systems, such as the common fcc (Fm-3m symmetry), bcc (Im-3m symmetry) or hcp (P63/mmc symmetry) metals, Eq 1. reduces to the commonly referred to "single-lattice" model ($S_{config}^{single-lattice} = -R \sum_i X_i \ln X_i$) due to the number of Wykoff positions (atom positions) within the unit cells. High entropy ceramics[7] are more complex: ceramics have separate cation and anion sublattices, and in high entropy ceramics at least one cation sublattice is configurationally disordered while the anion sublattice usually remains ordered. High entropy ceramics based on structures with multiple cation sublattices (e.g. perovskites) can have both ordered and disordered cation sublattices. Under these conditions, Eq. 1 should be used. In the oxide literature, this is sometimes referred to as using a "sub-lattice" model. Note that the expressions for configurational entropy for both models are based on the thermodynamic assumptions designed for alloys and are not specifically designed to differentiate between different oxides. For high entropy alloys, it is assumed that particular materials are defined purely by the value of $S_{config}$ described by Eq. (1) with classifications of high entropy ( > 1.5R), medium entropy (1 –1.5R), and low entropy ( < 1R). Many authors apply this division also to oxides and use the same relation to calculate $S_{config}$ as given for alloys[33] (Figure 2).

The single lattice ideal entropy approximation in Eq. (1) is only valid in the fully disordered, high-temperature limit, when all of the possible configurations are thermodynamically accessible. Many multi-component systems will feature some level of ordering, particularly short-range ordering, that will significantly reduce the entropy below the ideal value. The difference between the ideal entropy and the actual entropy of the material at finite temperature is called the relative entropy (Kullback-Leibler divergence[34]), which represents the entropy loss due to ordering.

When comparing configuration entropies among different systems, units need to be kept in mind, specifically whether the configuration entropy is in a per mol basis, per cation basis or per atom basis. Consider a 5-component metal (5-M) and oxide (5-MO). For 5-M, the configurational entropy on a per cation basis will be 1.609R and per atom basis will be 1.609R. However, for 5-MO, configurational entropy on a per cation basis will be 1.609R (comparable to the metal) but on a per atom basis it will be 0.805R (less when compared to the metal).

In the strictest sense, a system is entropy stabilized when the enthalpy of formation ($\Delta H_f$) is positive, and the entropy of formation ($\Delta S_f$) is positive and large enough to make the Gibbs free energy of formation ($\Delta G_f = \Delta H_f - T\Delta S_f$) negative above some temperature (T). The entropy of formation is typically broken into two contributions: $\Delta S_f = \Delta S_{Thermal} + \Delta S_{config}$. If the components of a compound are ordered, ($S_{config} = 0$), and the crystalline compound is perfectly disordered (Eq. 1), then the $\Delta S_{config} = [-R\sum_j m_j \sum_i X_{ij} \ln X_{ij} - 0]$. As the number of components grow, $\Delta S_{config}$ will increase if the system maintains perfect disorder. That is, if the additional component is non-interacting and follows the ideal mixing assumption. However, in phase equilibria it is not $\Delta S_{config}$ that determines phase stability, but rather the partial molar entropy which is defined as $\frac{dS}{dn_i} = R\ln X_i$, under the ideal mixing assumption[35]. This is because the partial molar entropy is directly related to chemical potential $\left(\mu_i = \frac{dG}{dn_i}\right)$ and it is the chemical potential of the components in a solution that controls solubility limits and phase stability.

When considering mixing in a multi-component system, it is easier to use the activity ($a_i$) of a new component, *i* being added to solution. This solution can be a pre-existing solution with multiple components, or a pure phase. The activity of a component *i* in solution is given by Eq. 2:

(2) $a_i = \gamma_i X_i$

where $X_i$ is the mole fraction of species *i* and $\gamma_i$ is the activity coefficient. The activity coefficient is a parameter that describes deviations from ideal mixing with the addition of a new component *i* to the solution. The activity of a species *i* in the solution is directly related to the change in chemical potential of species *i* in the solution in Eq. 3:

(3) $\Delta\mu_i = RT \ln a_i$

The true beauty of this relationship is shown when combining Eq. 2 and Eq. 3, giving Eq. 4:

(4) $RT \ln a_i = RT \ln \gamma_i + RT \ln X_i$

This relationship allows the activity of component *i* in a solution to be naturally broken into two separate contributions: *(i) an ideal mixing contribution ($RT \ln X_i$) and (ii) a non-ideal mixing contribution ($RT \ln \gamma_i$)*. This ideal mixing contribution is the same as the partial molar entropy derived from ideal, perfect mixing and is directly related to the phase stability, as described above. This is the portion of the configurational entropy that works to effect phase stability under the ideal mixing assumption. Notice how this contribution is independent of the number components. If mixing is ideal, then the non-ideal mixing component can be ignored (i.e. $RT \ln \gamma_i = 0$). However, it is this non-ideal mixing term that contains all the interesting effects of mixing in a solution. $RT \ln \gamma_i$ is referred to as the excess partial free energy or the excess chemical potential and contains effects such as: the heat of mixing $(\Delta H_i^{mix})$, the thermal entropy of mixing $(T\Delta S_{i,thermal}^{mix})$, the free energy of transformation between structures $(\Delta G_i^{\alpha \to \beta} = \Delta H_i^{\alpha \to \beta} - T\Delta S_{i,thermal}^{\alpha \to \beta})$, interface effects $(\Delta G_i^{interface} = \Delta H_i^{interface} - T\Delta S_{i,thermal}^{interface})$, magnetic effects $(\Delta G_i^{magnetic} = \Delta H_i^{magnetic} - T\Delta S_{i,thermal}^{magnetic})$, electronic effects $(\Delta G_i^{electronic} = \Delta H_i^{electronic} - T\Delta S_{i,thermal}^{electronic})$, defect effects $(\Delta G_i^{defects} = \Delta H_i^{defects} - T\Delta S_{i,thermal}^{defects})$, etc (Eq.8). It is this non-ideal contribution that gives rise to excess properties that deviate from a linear combination of their components and give rise to the so-called "cocktail effect".

Not all of the above effects will be active in a given system and will depend on the types of components that are being mixed. Because oxide solid solutions generally have a larger contribution of ionic and covalent bonding than metallic alloys, the importance of these various factors in determining mixing properties of oxides can be quite different. One important point in addressing synthesis of CCOs and HEOs is the solubility limit, which will be determined by the largest non-ideal mixing factor. In oxides, this is most commonly from the transformation energy due to its size compared to others. However, on the nano-scale, even the surface energy contribution becomes large enough to shift stable polymorphs with examples including $TiO_2$[36–38], $Al_2O_3$[39–42], $ZrO_2$[43–45] and $HfO_2$[44], and even solvus lines in multi-phase systems, such as CoO-ZnO[46,47]. To further emphasize the role of the non-ideal mixing term, it is illuminating to realize that systems can be entropy stabilized independent of the number of components. There is direct evidence for this in a series of oxides with components far less than 5, where the enthalpy of formation is positive and must be overcome by the entropy of formation for the system to be stable. Examples include certain two-component copper spinel's ($CuAl_2O_4$, $CuCr_2O_4$, $CuFe_2O_4$, $CuGa_2O_4$, $CuMn_2O_4$ and others)[48–51], (ii) pseudobrookites[52], sheared $TiO_2$-$Nb_2O_5$ structures[53] and the $A_6B_2O_{17}$ (A = Zr, Hf and B = Nb, Ta) family[54,55].

Disordered materials can be modeled using special quasi-random structures (SQS)[56], in which the lattice is decorated so as to maximize the similarity of the radial correlation functions to that of the completely disordered material. SQS models the fully-disordered, infinite-temperature limit, which is useful for understanding how high levels of disorder can affect properties, but it neglects the effect of short- and long-range ordering at finite temperatures. Alternatively, disorder can be represented by an ensemble of small ordered cells, such as in cluster expansion[57] or AFLOW-POCC[58]. These small cells represent the different configurations that form "tiles", which are randomly arranged to generate the disordered material. By thermodynamically weighting the properties of the different tiles, the behavior of the disordered material can be modeled from the low temperature state, which has a high degree of ordering due to only a few low-energy configurations being present, up to the high-temperature state where all configurations have a similar probability of being present. Descriptors based on the distribution of energies of the configurations can also be used to predict the synthesizability of single phases[59]: narrow distributions imply that it is easy to introduce new configurations to increase disorder, while broad ones indicate preferences for specific ordered states. The complexity of high entropy systems makes them computationally challenging to model directly, making machine-learning an important tool for future investigations of such systems, both in the form of training interatomic potentials[60] as well as for directly predicting properties and synthesizability[61].

Based on mixing principles and evidence in the literature, it is clear that the non-ideal mixing parameters are important and need to be taken into consideration when trying to design new and exciting entropy stabilized oxides. Central to this is the ability to control the degree of mixing, achieving an ideal mixed state, and subsequently characterizing the resulting materials experimentally.

*Experimental Control of Mixing through Synthesis*

Since its advent, the multicomponent approach has opened a broad space for compositional design with many new materials being synthesized in various polycrystalline, single crystal, and nanocrystalline forms. While synthesis is at the root of all experimental materials discovery and characterization, the speed at which this field has advanced with new compositions and investigations readily coming out in the literature has not allowed for full in-depth considerations of the effects of synthesis method on the materials, their phase formations, and their reported properties. However, it is clear that the method of synthesis combined with thermal post-processing can be used to control the degree of compositional mixing within a material.

In the case of the seminal rocksalt $(Mg_{0.2}Ni_{0.2}Co_{0.2}Cu_{0.2}Zn_{0.2})O$, its entropy stabilized phase is well documented[62]. However, the disorder within the system and specifically the distortions for the copper atoms can drive modifications to crystal structure and subsequently functionality through post-synthesis heat treatments[63]. This brings an interesting point to consider when material properties are reported, as the method and temperature by which a sample is prepared could have effects on the properties, especially for those strongly dependent on smaller length scale interactions. There have been a few studies that discuss the role of temperature and have confirmed that not only configurational entropy but also the temperature-time profile both play a role in single phase formation[64]. In-situ studies to evaluate the kinetic mechanisms involved in the phase formation of polycrystalline multicomponent oxides showed both expected and unexpected differences in the mechanisms seen both across synthesis method employed and compositions containing, indicating the underlying contributors that need to be considered[65].

| Polycrystalline | Single Crystal |
| --- | --- |
| •solid state methods (ambient, high-pressure) | •pulsed laser deposition |
| •sol-gel | •micro-pull-down |
| •nebulized spray pyrolysis | •floating zone |
| •flame spray pyrolysis | |
| •steric entrapment | |
| •co-precipitation | |
| •solution combustion synthesis | |
| •spark plasma sintering | |
| •hydrothermal | |
| •mechanochemical | |

Table 1: Material synthesis approaches utilized for compositionally complex multicomponent materials.

It is known and expected that there will be differences in material properties between single crystal and polycrystalline materials of a given composition, but the inherent disorder and the kinetic mechanisms' contributions to ordering in phase formation among a given crystalline form across synthesis methods and temperatures could become a more crucial factor. This is critical when properties are closely tied to local ordering. Specifically, understanding this detailed relationship means that careful attention to the length scale of disorder is an important aspect. It has been seen in a few varieties of studies that such dependencies on synthesis and elemental composition can have effects, but not in all cases[66], which further strengthens the need for such investigations.

A variety of synthesis techniques, listed in Table 1, have been employed in the formation of multicomponent oxide materials. It has been seen that single-phase stabilization can often be dependent on the method of synthesis employed. Non-equilibrium growth methods, such as pulsed laser deposition (PLD), have proven successful at stabilizing single crystals with random mixing on the desired sublattice[32]. This is likely a response to the extraordinarily short quench times associated with laser ablation which maximizes mixing during synthesis. Equilibrium bulk growth techniques such as floating zone and micro-pull-down are able to generate single crystals but with slight elemental gradients within the composition[67]. Furthermore, multicomponent polycrystals that are unable to achieve single phase can be used as a target precursor for non-equilibrium growth methods. The secondary phases seen in the bulk are not observed in the grown single crystal. Spray pyrolysis and other nano-particle synthesis techniques have been shown to result in single phase particles but are likely to have surface and geometry driven segregations that also need to be considered. This again ties into the local structure discussion and the need to probe across various length scales in order to truly understand mixing and whether a particular composition and sample is entropy stabilized.

Reversibility in crystallographic phases has been reported and studied within a few classes of multicomponent oxides[62,68–70]. However, longer term studies on phase stability are not yet available. Some multicomponent oxides have shown high water incorporation into the materials structure leading to good proton conductivity properties [71]. Water absorption leads to the question of shelf life, especially in humid environments, and is an important property to further explore especially for those materials that hold promise for use in catalysis [72] and energy storage [73].

*Characterization of Crystallinity, Composition, and Mixing at All Length Scales*

As introduced above, a random and homogenous cation distribution on a single sublattice site with no evidence of chemical clustering is often considered a defining characteristic of HEOs[74,75]. When multiple ion types are found on single crystallographic sites, the intrinsic variation in ionic size, charge, or spin state may impart chemical short-range order (CSRO) or compositional patterning on the sublattice. Many physical properties of complex oxides are governed not just by the identity and local ordering of participating cations but also by the physical length-scale characteristic to the patterning. The occurrence of multiple, different nanostructured regions within a single chemical composition have been found advantageous for enhancing properties in thermoelectric materials, radiation tolerant ceramics, and relaxor ferroelectrics, among many other natural and engineered systems[76,77]. Random and homogenous distributions may well prove to be a rare occurrence among compositionally complex ceramics. Additionally, anti-site defects, non-stoichiometry, and all manner of local atomic distortions may prove to dominate their complex and richly variable crystal chemistries. Thus, configurational complexity (CSRO, local atomic distortions, vacancies, etc.) and the length scale over which it extends within a material is equally important in evaluating phase stability and understanding physical properties in this arena. It was indicated above that synthesis and processing conditions can offer more extensive opportunities to control specific cation order, defects, and distortions present in compositionally complex materials, offering expansive opportunities for material design guided by careful characterization of the differences of both long- and short-range order in these systems.

Classification of CCMs requires characterization across all examinable length scales (Fig. 3). As with HEAs, there are a variety of techniques in the literature[78] that have been used to understand the degree of spatial variance in specific compositions in terms of compositional mixing at sub-nanometer length scales and local deviations from global lattice symmetries. The characterization path in which an investigator chooses to take depends on several factors including technique availability and the specific questions needing clarity. With complex ceramics, particularly those classified as high or medium entropy, a suite of structural and compositional characterization techniques is needed to gain a holistic picture of these systems and how they might fall into the aforementioned classes of high entropy, medium entropy, and entropy stabilized materials. High entropy materials should maintain a homogeneous distribution of elements across all length scales, whereas medium entropy systems may have some level of segregation across differing length scales, but the segregation itself maintains already established entropy definitions. It is important to note however, that structural and compositional characterization alone are not sufficient to define an entropy stabilized material: calorimetry-based measurements must also show evidence of endothermic transition[79,80].

X-ray diffraction (XRD) is perhaps the most widely used characterization technique, as it provides rapid information on macroscopic phase purity, crystal symmetries, and lattice structure. Advanced analyses such as Rietveld refinement[81] can shed light on peak asymmetries, broadening, and missing (or extra) reflections whether that be from crystallite size and strain, or ordering within lattice planes. For example, Usharani, *et al.*[82] determined that a systematically increasing asymmetry along all Bragg peaks in high entropy rocksalt $(MgCoNiCuZn)_{0.2}O$ (space group: Fm-3m) grown using different synthesis techniques to be caused by distortions leading to the formation of monoclinic (space group C1 2/m 1) CoO and NiO. Bérardan, *et al.* determined that the incorporation of Li into $(MgCoNiCuZn)_{0.2}O$ was substitutional, but the resulting decrease in lattice parameter contradicted an addition of a $1^+$ cation to the matrix leading to a hypothesis of charge compensation elsewhere in the system[83]. This was confirmed using X-ray photoelectron spectroscopy (XPS).

The shortfall of only using XRD comes from the similarities in scattering factor, absorption, etc. of the elements typically found in HEOs, and it becomes virtually impossible to extract local or short-range ordering. Even in cases where advanced XRD analysis is performed, additional complementary techniques are needed to strengthen confidence. Scanning and/or transmission electron microscopies equipped with spectrometers, energy dispersive X-ray (EDXS) or electron energy loss (EELS), shed light on elemental distributions from macroscopic down to nm (columns of atoms) length scales[23]. Atom probe tomography (APT) takes this a step further by creating a 3D atomic distribution map. Chellali, et al.[84] used APT to investigate atomic level homogeneity in a high entropy rocksalt, fluorite, and perovskite structure. Using depth profiling, they found each composition to be homogeneous throughout the entire sample.

Total scattering experiments, with neutrons and X-rays, can be used to understand atomic composition through macroscopic details. A recent study[85] on rare-earth pyrochlore oxide $Nd_2(TiNbSnHfZr)_{0.4}O_{7+x}$ indicated a phase pure material in XRD and STEM-EDS provided evidence of a homogenous cation distribution on the nm length scale. However, neutron total scattering measurements indicated that this model of homogeneity breaks down on the local scale of nearest neighbor distances, suggesting distinctive or ordered chemical environments. X-ray absorption fine structure (XAFS) complements total scattering in that it provides element-specific average local radial distributions and neighboring species. An initial XAFS study of $(MgCoNiCuZn)_{0.2}O$ investigated the local structure around four of the five cations, observing Jahn-Teller-type distortions of the $CuO_6$ octahedra, and large static disorder contributions to the EXAFS Debye-Waller factors. Fitting to next nearest neighbors (NNN) suggested a homogeneous distribution of cations on the smallest possible length scale, as fits containing clustering of more than three elements in the NNN shell resulted in poor fit statistics and/or non-physical results. Additional spectroscopy techniques including Fourier transform infrared spectroscopy (FTIR)[86] and Raman[87] have been used to explore local structure around specific elements in CCMs.

*Outlook*

Intuition reveals that many of the most interesting aspects and applications of "HEOs" and related materials will ultimately involve departure from ideal and undistorted lattice configurations. However, this introduces a problem in defining the class: to what extent must the atomic ordering motif be random/homogenous to qualify as having "high," "medium," or just "ordinary" configurational entropy? Taken to its extreme, will we need to know the identity and location of *every* atom in a compositionally complex oxide to exactly describe the various entropic and enthalpic contributions to its thermodynamic state and classify it? Or, is there an underlying selection of length-scale for averaging that can be identified? While there are a variety of options for experimentally distinguishing a true ESO from the wider class of HEOs, defining a set of criteria that can separate medium or high entropy oxides apart from compositionally complex oxides is far more difficult. Even if a succinct definition is agreed upon, there are significant characterization challenges associated with determining atomic and nanoscale order/disorder. Of the dozen or so studies investigating cation arrangement and/or short-range order (or in some cases spin disorder), nearly all have focused on the inaugural class of rock-salt HEOs [79,88,89]. The influence of local atomic structure, lattice distortions, and their impacts on physical properties, remains largely unexplored in HEOs to date [85,90] and more work is called for.

As the search for novel materials advances in this field, the diversity of crystal structures and compositional combinations will continue to increase (fig. 4). Rocksalts, spinels, and perovskites have thus far led the way; however new crystal phases such as $A_2B_2O_7$ pyrochlores, which offer unique structural frameworks, phase stability, and compositional flexibility are creating attractive playgrounds for researchers looking for exotic magnetism, enhanced radiation resistance, and superior thermal stability and ionic/electric conductivity[91–94]. Classifying these types of new systems with the entropy metrics described above will be important as we move toward finding commonality between classes. For example, a wide range of *A* and *B* cations can accommodate the pyrochlore structure, hence entropy is not expected to be the driving force for stability in most cases though exceptions are certainly expected[95]. Rare earth zirconates were used as a case study that demonstrated the ability to form a single phase $RE_2Zr_2O_7$ compound stabilized by the difference in cationic radius rather than the ideal entropy value[96]. In addition, it was shown that the disorder on the *A*-site of pyrochlores only affects the six nearest neighbor *B*-site cations, compared to the 12 nearest neighboring cations in the rock salt structure, thus reducing the configurational entropy in pyrochlores[97]. The "complexity" provided by mixing cations on two separate cation sites leaves room for additional subjectivity, as these compounds can have an overall mixing entropy that is higher than compounds with cation-mixing on a single cation sublattice but low mixing entropy on a single cation sublattice[98].

Moving beyond oxides, the considerations outlined above should continue to hold and allow the design of high-entropy stabilized chalcogenides[99]. Functionally, these systems offer extraordinary promise, as the low-complexity parent compounds are well known for their thermoelectric[99, 100], electrocatalytic[101], electrode[102], and superconducting[103] applications. Tuning sublattice complexity in these structures is showing new routes to manipulate thermal conductivity[100], redox potential[101], magnetic order[104], and superconductivity[103,104]. While difficult to synthesize, various synthetic methods have been developed[105,106]. As a next step, investigation of local crystal structure by PDF/TEM and local charge distribution by the XAS/XPS (e.g., a possibility of charge transfer between $Fe^{2+d}$ and $Cu^{2-d}$ in $(Fe,Co,Ni,Cu,Ru)S_2$ and $(Fe,Co,Ni,Cu,Ru)Se_2$ pyrite) should prove interesting counter points in comparison with those in HEOs including common elements. To understand the whole picture of HE ionic/covalent compounds, a comparison between HEOs and HE chalcogenides is inevitable; and points to many opportunities for new studies related to controlled variation of the anion sublattice.

*Concluding Remarks*

The terminology "compositionally complex" is typically applied to distinguish the presence of an intentional high number of cations substituted on a single sublattice site. This term only addresses the number of atom/ion types involved in the compounds, carrying no assumption regarding their random/homogenous distribution. At the other end, the strictest definition is that of "entropy-stabilized oxides", which refers to compositions in this set whose single-phase structures are stabilized through their high configurational entropy. This is a thermodynamic definition, and there are a variety of experimental validations to test for it. Somewhere in between lie the "medium" and "high entropy" oxides, which we propose may be distinguished by both the number of atoms involved (compositional complexity) and the degree of mixing on individual sublattices (configurational complexity). As a result, proving something is "medium" or "high" entropy may prove to be rather difficult, requiring an extremely rigorous characterization protocol and a comprehensive understanding of the thermodynamic factors at play. Should an agreed upon length-scale for homogeneity (and experimental and theoretical characterization protocols) be established alongside the compositional complexity threshold for high entropy materials? Should the high entropy labels be reserved to signify select cases where the compositional and configurational (local-to-long-range) complexity impart synergistic effects or emergent properties to materials? Regardless of the definitions pertaining to these materials, untangling complexity at all length-scales of interest may ultimately prove necessary for finding the keys to their stability and evaluating how extensively their functionality may be varied and controlled.

**Acknowledgment**

This work is the result of discussions occurring at the Telluride Workshop on Compositionally Complex Oxides. All authors contributed equally. Work at Oak Ridge National Laboratory was supported by the US Department of Energy (DOE), Office of Basic Energy Sciences (BES), Materials Sciences and Engineering Division. The work at Los Alamos National Laboratory was supported by the NNSA's Laboratory Directed Research and Development

Program, and was performed, in part, at the CINT, an Office of Science User Facility operated for the U.S. Department of Energy Office of Science through the Los Alamos National Laboratory. Los Alamos National Laboratory is operated by Triad National Security, LLC, for the National Nuclear Security Administration of U.S. Department of Energy (Contract No. 89233218CNA000001). The National Science Centre, Poland, project number 2019/35/B/ST5/00888 (MG, AMG). KAKENHI, Grant-in-Aid on Innovative Areas (#18H05462), Scientific Research(A) (#18H03692), and Scientific Research(C) (#20K05450), GRIMT program (#202012-RDKGE-0065) at IMR, Tohoku University (AY). C.M.R. gratefully acknowledges the support from NSF through the Materials Research Science and Engineering Center DMR 2011839. SJM acknowledges support from the National Science Foundation (NSF), Division of Materials Research (DMR), Ceramic (CER) program under grant DMR 2047084

**AUTHOR DECLARATIONS**

**Conflict of Interest**

The authors have no conflicts to disclose.

**DATA AVAILABILITY**

The data that support the findings of this study are available from the corresponding author upon reasonable request.

**Figure Captions:**

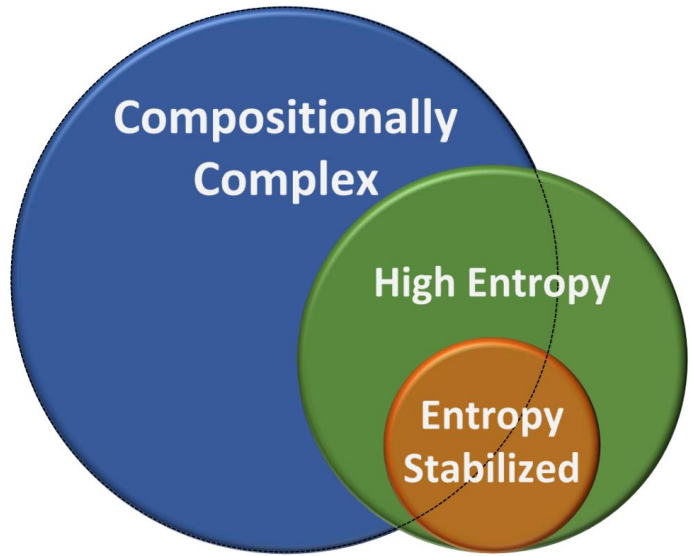

Figure 1. Hierarchy of terminology commonly used to define materials having compositional disorder on an ordered lattice. All entropy stabilized materials are high entropy materials; however not all high entropy materials are compositionally complex.

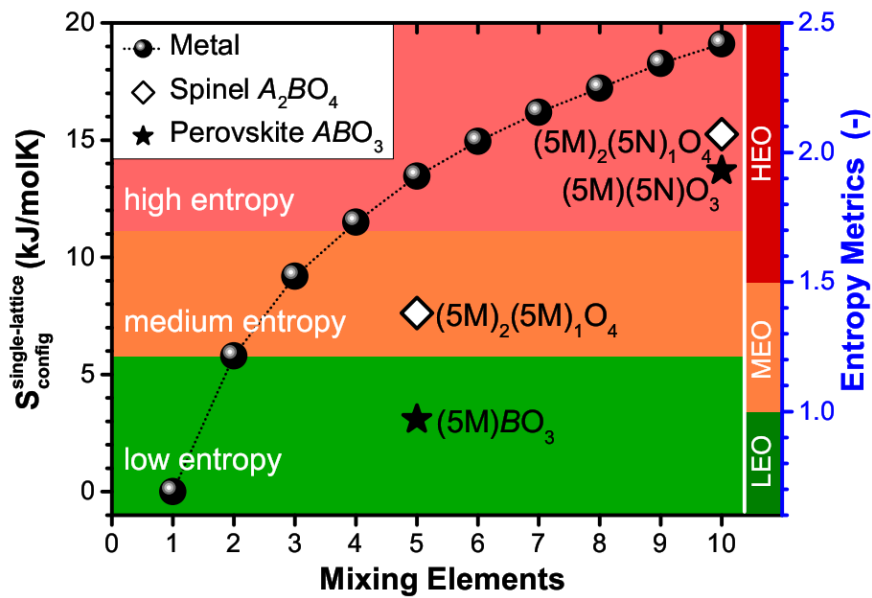

Figure 2. Visualization of difference in entropy evaluation by single lattice model used for metals and entropy metrics used for complex oxides[33].

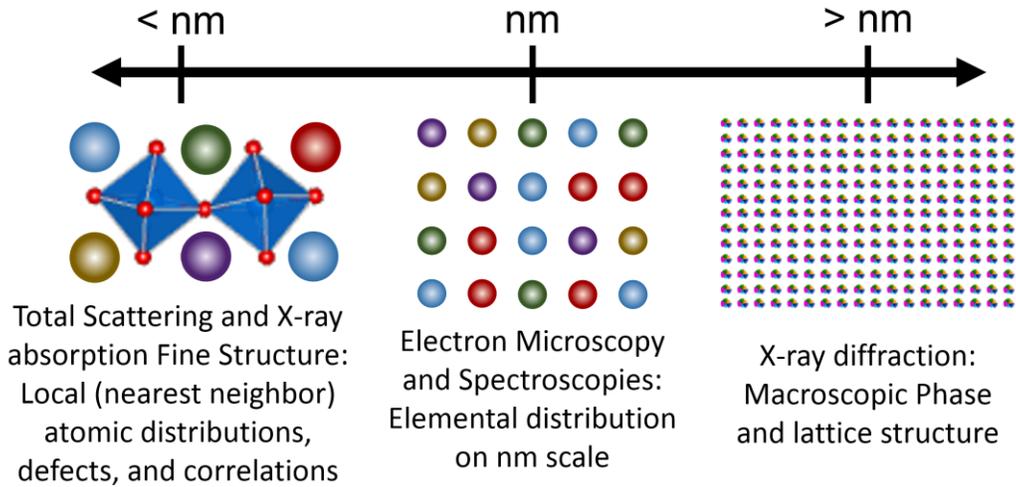

Figure 3. Defining disorder across length scales requires a range of characterization techniques to understand the type and scale of entropy contributors.

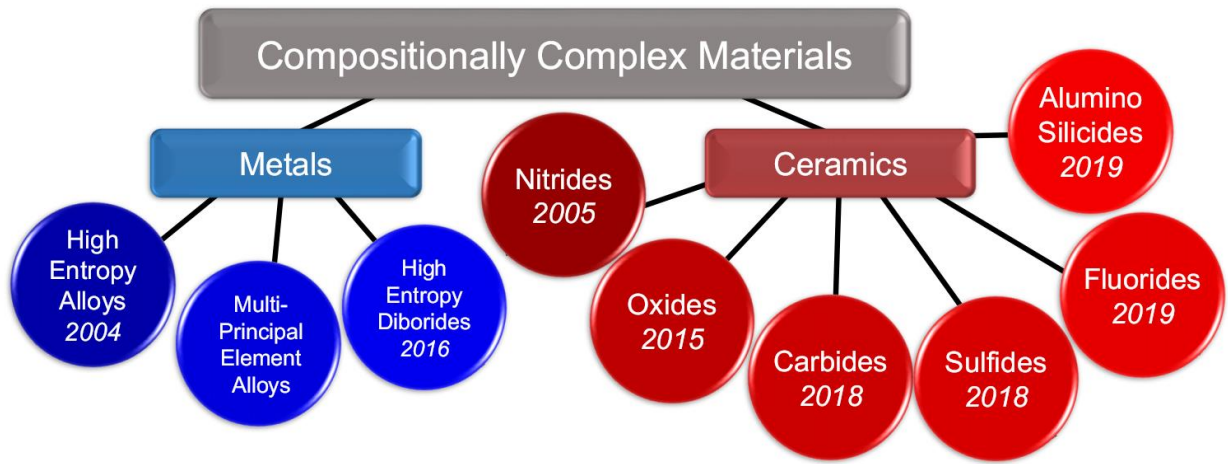

Figure 4. Application of compositional complexity continues to grow as a novel approach to design new functionalities into an ever-broader range of materials classes.